# Infant Cry Detection In Noisy Environment Using Blueprint Separable Convolutions and Time-Frequency Recurrent Neural Network

Haolin Yu
School of Electronic and Information Engineering
South China University of Technology
Guangzhou, China
202230244210@mail.scut.edu.cn

Yanxiong Li*
School of Electronic and Information Engineering
South China University of Technology
Guangzhou, China
eeyxli@scut.edu.cn

*Abstract*—Infant cry detection is a crucial component of baby care system. In this paper, we propose a lightweight and robust method for infant cry detection. The method leverages blueprint separable convolutions to reduce computational complexity, and a time-frequency recurrent neural network for adaptive denoising. The overall framework of the method is structured as a multi-scale convolutional recurrent neural network，which is enhanced by efficient spatial attention mechanism and contrast-aware channel attention module, and acquire local and global information from the input feature of log Mel-spectrogram. Multiple public datasets are adopted to create a diverse and representative dataset, and environmental corruption techniques are used to generate the noisy samples encountered in real-world scenarios. Results show that our method exceeds many state-of-the-art methods in accuracy, F1-score, and complexity under various signal-to-noise ratio conditions. The code is at https://github.com/fhfjsd1/ICD_MMSP.

## I. Introduction

With the acceleration of modern life pace, the cost for infant care is constantly increasing. As one of the primary means for infants to communicate with the outside world, crying provides critical insight into the infants' physiological state and often indicates abnormal conditions. Hence, accurate and real-time detection of infant cries not only helps reduce parenting risks at home but also offers extensive application prospects in intelligent infant care systems.

In recent years, advanced signal processing techniques and data driven algorithms have demonstrated significant potential in the study of infant cries, gradually enabling detection [1-4], classification of the underlying causes [7-13], and even pathological diagnosis [14-16]. Despite this promising performance, most studies have been conducted on single, controlled, small scale datasets. Even the widely recognized BABY CHILLANTO dataset [17] has only one recording for each of 127 newborns. The reliance on uniform recording conditions and a limited number of subjects may introduce significant homogeneity, which in turn leads to algorithmic overfitting. Moreover, many researchers have employed datasets that consist of laboratory-processed, clean recordings or that introduce only a limited and discrete selection of noise events as negative samples [1-3], resulting in data that lack representativeness of real-world scenes. Although other datasets [4-6] have been collected in home environments, their detailed descriptions and quantitative metrics regarding noise factors are unavailable. The absence of data diversity and representativeness of noisy scenarios further implies that these methods may underperform when deployed in real world.

Traditional machine learning approaches that rely on handcrafted features fail to capture the variable patterns of infant cries and are sensitive to environmental noise. Deep learning techniques have shown superior performance [1, 3, 4]. However, many existing studies have limited themselves to basic convolutional neural network (CNN), recurrent neural network (RNN), and their variants. These models often exhibit suboptimal performance on complex datasets and lack the elements tailored to the unique attributes of infant cries. Although some works [7-9] have been made on sophisticated models to improve performance, their computational demands and energy consumption hinder their practical deployment to resource-constrained devices.

In summary, there remains a critical need for methods that can effectively detect infant cries in the context of household noise environments while offering practical applicability [14]. In this work, we make the following three contributions.

1) We propose a data integration strategy that combines multiple public datasets to create a diverse and cross-scenario representative dataset for infant cry detection, along with a data processing strategy tailored for real-world applications.

2) We design a low-complexity model which consists of multi-scale Blueprint Separable Convolutions (BSConv) and RNNs. The BSConv module is augmented with attention mechanism, while the RNN establishes dependency in both time and frequency domains to denoise adaptively.

3) We propose a method for infant cry detection in noisy environment using blueprint separable convolutions and time-frequency recurrent neural network. Experimental results prove the effectiveness and efficiency of the proposed method.

## II. Related work

In recent years, numerous audio-based studies have investigated various feature-engineering techniques in conjunction with traditional machine learning and other algorithmic frameworks for cry detection. Lee Sze Foo et al. [5] employed audio features (such as pitch, zero-crossing rate) to detect cries, and obtained an accuracy of 89.2%. Liu et al. [10] extracted audio features, such as linear predictive coding,

*Corresponding author: Yanxiong Li (eeyxli@scut.edu.cn).

This work was partly supported by the national natural science foundation of China (62371195, 62111530145, 61771200), exchange project of the 10th meeting of the China-Croatia science and technology cooperation committee (10-34), provincial undergraduate training program for innovation and entrepreneurship (S202410561216), and Guangdong provincial key laboratory of human digital twin (2022B1212010004).

bark frequency cepstral coefficients, and Mel-frequency cepstral coefficients, and used the compressed sensing theory to classify infant cries. Moreover, traditional classifiers, including random forest, logistic regression, support vector machine (SVM), and decision tree, were utilized to make decisions on time-domain features extracted from the autocorrelation function [3] or frequency-domain features such as bandwidth and spectral roll-off [13].

With the advent of end to end approaches, deep learning techniques have increasingly been applied to the analysis of infant cries. For instance, Yao et al. [4] utilized a CNN to extract deep features that were fed to a SVM classifier with a radial basis function kernel. Anjali et al. [11] fine-tuned the memory intensive VGG16 model on spectrograms and achieved an accuracy of 0.92 on the Dunstan Baby Language dataset. Long Short-Term Memory (LSTM) network has been employed for spectrogram analysis as well [1]. Additionally, Shen et al. [7] simultaneously utilized both CNNs and RNNs for cry recognition. Similarly, Zhang et al. [8] leveraged a sparse autoencoder LSTM and VGGs for deep feature extraction and fusion. Further studies include the work of Xia et al. [2], who proposed a multi-task learning method for audio-based infant cry detection and reasoning.

## III. METHOD

### A. Preliminary

*1) Blueprint Separable Convolutions*

The BSConv [18, 19] utilizes intra-kernel correlations and allows for a more efficient separation of regular convolutions. Specifically, BSConv decompose a standard convolution into a pointwise convolution along with a depthwise convolution. Consider a standard convolutional operation employing $N$ filters, $F^{(1)}, \ldots, F^{(N)}$, each of size $C \times K \times K$, where $C$ denotes the depth dimension of the input tensor and $K$ denotes the kernel size. The total number of free parameters in this operation is

$$(C \times K^2) \times N. \quad (1)$$

In contrast, BSConv re-parameterizes the $n^{th}$ filter $F_{c,:,:}^{(n)}$ of its $c^{th}$ channel as

$$F_{c,:,:}^{(n)} = \omega_c^{(n)} \cdot B^{(n)}, \quad (2)$$

where $\omega_c^{(n)} \in \mathbb{R}$ denotes pointwise convolution weight, and $B^{(n)}$ denotes a 2D blueprint used in the depthwise convolution. Therefore, the number of parameters is reduced to

$$(C + K^2) \times N. \quad (3)$$

*2) Enhanced Spatial Attention*

The block of Enhanced Spatial Attention (ESA) [20] represents a lightweight and efficient attention module. As shown in Fig. 1, convolutions and max-pooling are first applied to reduce the channel and spatial dimensions. Subsequently, convolution groups, implemented by BSConv as well, are employed to extract features, and bilinear interpolation-based upsampling are used to restore the original spatial resolution. Finally, the resulting attention mask is then multiplied element-wise with the original input features.

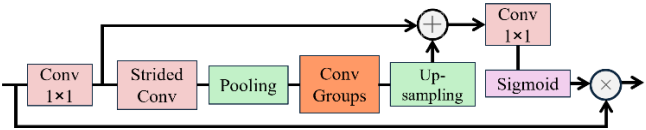

Fig. 1 *Framework of enhanced spatial attention block.*

*3) Contrast-aware Channel Attention*

The module of Contrast-aware Channel Attention (CCA) leverages the sum of the standard deviation and the mean (contrast of the feature map) to adjust the weight for each channel. The dimension of an input tensor $X$ is $C \times H \times W$, where $C$, $H$ and $W$ denote the number of channels, height, and width of the feature maps, respectively. Let $Z_c$ denote the global contrast for the $c^{th}$ channel. The contrast information is computed by

$$Z_c = \sqrt{\frac{1}{H \times W} \sum_{(i,j) \in X_{c,:,:}} (X_{c,i,j} - m)^2} + m, \quad (4)$$

where $m$ denotes the mean of a channel and is calculated by

$$m = \frac{1}{H \times W} \sum_{(i,j) \in X_{c,:,:}} X_{c,i,j}. \quad (5)$$

### B. Proposed Model

Models based on autoencoder are effective in denoising tasks [21, 22]. Inspired by these findings, we propose a model that processes audio signals in the Time-Frequency (T-F) domain. The proposed model comprises three components, including an encoder, an Adaptive Denoising Module (ADM) and a classifier (omitting the decoder), as shown in Fig. 2. Mel-spectrogram of each sample is fed to the encoder as input for extracting embedding which is further transformed by the ADM. The ADM primarily relies on RNNs to extract global features, and suppresses potential noise by establishing long-term dependencies in the T-F domain. The classifier, consisting of a multi-layer perceptron, projects the high-dimensional features to produce the final classification outcome. The details of each module are described below.

*1) Encoder*

The encoder is responsible for compressing information and extracting local information of the input feature. To reduce computational overhead, we replace standard convolutions with BSConv. Multiple BSConv blocks together with pooling layers are stacked for embedding extraction. Deeper networks are generally able to enhance the semantic representation of feature maps. However, deeper networks may result in the loss of critical spatial geometric details in the spectrograms of infant cries, especially those related to the fundamental frequency and its harmonics. To mitigate the problem, akin to skip connections, we concatenate feature maps obtained at different scales along the channel dimension to balance abstract semantics with geometric detail. Given that human voice spectrograms exhibit distinct texture patterns and dynamic intensity variations, we assume that channel attention based on contrast information can compress and fuse the concatenated features, further eliminate redundant information, distinguish noise, and enhance the expressive quality of infant cries. Moreover, to effectively leverage more representative features without significantly increasing network depth and complexity, the ESA mechanism is applied at the end of each block. As a result, the extracted embeddings are forced to concentrate on regions of interest.

*2) Adaptive Denoising Module*

Household noise is inherently complex and diverse, comprising transient high-intensity interferences, continuous white noise, and human vocalizations. It can be characterized by distinct T-F features. Within a short-time window, the temporal pattern of infant cries exhibits more pronounced quasi-stationarity compared to noise. In the frequency domain, infant cries display clear harmonic structures intrinsic to human vocalizations. Therefore, compared with conventional approaches that model only temporal dependencies [23], our method leverages the local information captured by CNNs and employs RNNs to model global dependencies between consecutive frames in both the time and frequency domains. Specifically, we treat the frames obtained from the Short Time Fourier Transform (STFT) as processing units. Within each unit, we apply a Bi-directional LSTM (BiLSTM) to model the spectral pattern of individual frames. Therefore, long-term harmonic correlations can be captured. Then, an LSTM is used to learn temporal dependencies at specific frequency bins. Additionally, a residual connection is applied between the input and output of the LSTM and the BiLSTM to further mitigate the issue of gradient vanishing.

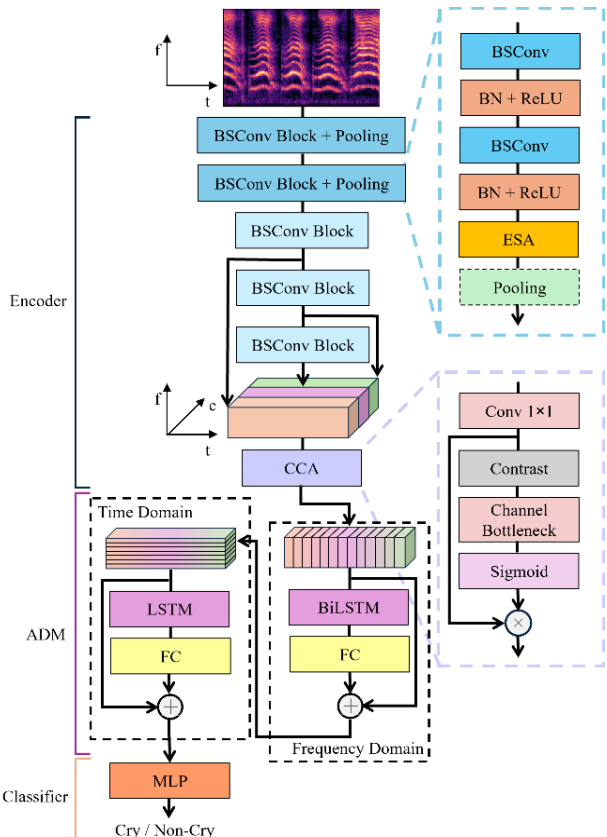

Fig. 2 *Framework of the proposed model. BN: Batch Normalization; FC: Fully Connected layer; MLP: Multi-Layer Perceptron; ReLU: Rectified Linear Unit; ESA: Enhanced Spatial Attention; ADM: Adaptive Denoising Module; LSTM: Long Short-Term Memory; BiLSTM: Bi-directional Long Short-Term Memory; CCA: Contrast-aware Channel Attention; BSConv: Blueprint Separable Convolutions.*

## IV. EXPERIMENTS

### A. Dataset

Public datasets of infant cries are extremely limited in both size and diversity of patterns. Consequently, experimental dataset employed in our study was constructed by integrating multiple public datasets to ensure that the samples are diverse and representative. The experimental dataset is approximately 39.6 hours of audio recordings in length. We retained only the audio files, and subsequently partitioned the recordings into two balanced classes: cry (positive) and non-cry (negative).

The cry data subset consists of audio recordings of infant cry derived from the CryCeleb2023 [24], EnesBabyCries1 [25] and 2020 iFLYTEK A.I. Developer Competition [26]. The datasets of VoxCeleb [27], ESC 50 [28], Cat Meowing [29] and DASEE [30] are used as non-cry data and further divided into five subclasses: Speech, Human non-speech, Cat Meows, Household noise, and Silences. Specifically, Cat Meows are notably confusable with cries as noted in [31]. Household noise encompassed sounds from domestic appliances, outdoor natural environments, and urban traffic, among others.

According to the previous study [32], the typical high vocal tract formant of an infant cry is usually below 6 kHz. Hence, all samples are resampled to a single-channel 16 kHz WAV PCM file with 5 seconds in length, adequately to captures one or multiple cycles of infant cry. In practice, infant monitoring devices can implement a sliding window with an energy threshold to trigger short-term detection, thereby maintaining low energy consumption.

For each cry sample, silent segments are removed since they generally do not contain useful acoustic information. The cry segments are then split or concatenated to form an audio sample. We finally obtain 13330 cry samples and each cry sample is of 5 seconds. Non-cry samples are generated in the same way and each sample is of 5 seconds. The number of samples of different classes are listed in Table I.

TABLE I NUMBERS OF SAMPLES OF DIFFERENT CLASSES.

| Cry | Speech | Household | Non-speech | Cat Meows | Silences |
|---|---|---|---|---|---|
| 13330 | 4963 | 4506 | 2076 | 1953 | 1650 |

For each processed sample, a Hanning window of size 512 is applied to segment the sample into frames with a hop length of 400. A Mel-scale STFT is computed using 128 triangular filter banks to derive the linear power spectrum. To achieve regional normalization, the triangular Mel weights are divided by the width of the Mel band, and a logarithmic transformation is subsequently applied to yield log Mel-spectrogram of 128 dimensions. The log Mel-spectrogram has been widely used as the input of deep neural networks for many types of audio processing tasks, such as audio clustering [33-36], speaker recognition [37-38], and audio classification [39-42]. Hence, it is also used in the study.

### B. Data Augmentation

Three data augmentation methods (speed perturbation, environmental corruption, time and frequency masking) are executed randomly during each iteration.

#### 1) Speed Perturbation

This method involves resampling the audio sample at a random rate similar to the original, resulting in a slightly slower or faster rendition of the signal. This operation uniformly selects a speed-up factor from 0.8 to 1.2 relative to the original sampling rate and also influences other acoustic features, such as pitch and formant frequencies [43].

#### 2) Environmental Corruption

In real-world scenarios, the signals are often contaminated by undesired noises. Consequently, beginning with a clean signal, various types of disruptions are introduced in a controlled manner.

For each pair consisting of a waveform vector $\hat{x} \in \mathbb{R}^L$ and a noise vector $\hat{n} \in \mathbb{R}^L$, additive noise is scaled according to a specified signal-to-noise ratio (SNR) and then added to the original waveform, following a defined formula:

$$\hat{y} = \hat{x} + a\,\hat{n}, \quad (6)$$

where $a$ is a coefficient and defined by

$$a = \sqrt{\frac{||\hat{x}||_2^2}{||\hat{n}||_2^2} \cdot 10^{-\frac{SNR}{10}}}. \quad (7)$$

In an acoustic environment such as rooms, multipath propagation caused by sound reflections, known as reverberation, can significantly degrade the clarity of signals. The reverberation effect between the source and the receiver is modeled by an impulse response, with which reverberation is introduced via convolution. The room impulse responses employed in this process are obtained from [44].

*3) Time and Frequency Masking*

This technique involves randomly replacing contiguous segments of the original sample with zero values in both the time and frequency domains. Such augmentation encourages the model to rely on information distributed throughout the entire sample rather than particular parts.

*C. Experimental Setup*

To validate the effectiveness of our proposed method, we compare our method with some state-of-the-art ones under various noise conditions characterized by different SNRs.

We choose the following models: BiLSTM, ResNet18 [45], ECAPA-TDNN [46], and MobileNetV2 [47]. The log Mel-spectrogram is used as the input of all models. Bilinear interpolation is used to resize the spectrograms appropriately to match required input size of each model. Experiments are conducted under the same conditions. The training process is run for up to 200 epochs with early stopping implemented. The Adam optimizer is employed in combination with a cyclical learning rate policy, which cyclically adjusts the maximum learning rate between specified lower and upper bounds over a cycle of 20,000 batches [48]. A binary cross-entropy loss is employed as the criterion. Multiple evaluation metrics are used, and five-fold cross-validation is performed with the average results reported.

*D. Experiment Results*

We initially train our models on a clean dataset and then evaluate them on the clean dataset and on the noisy data with SNR of 0 dB. Results are shown in Table II.

TABLE II EXPERIMENTAL RESULTS OBATINED BY DIFFERENT METHODS WHEN THEIR MODELS ARE TRAINED WITH CLEAN DATASET AND THEN TESTED ON CLEAN AND NOISY DATASETS (IN %).

| Methods | Metric | Clean | Noisy | Margins |
|---|---|---|---|---|
| BiLSTM | Accuracy | 96.9 | 66.8 | 31.1 |
| | Precision | 96.8 | 96.6 | **0.2** |
| | Recall | 96.7 | 30.2 | 68.8 |
| | F1-score | 96.8 | 46.0 | 52.4 |
| ECAPA-TDNN | Accuracy | 97.9 | 74.1 | 24.3 |
| | Precision | 98.1 | 92.1 | 6.1 |
| | Recall | 97.4 | 48.9 | 49.8 |
| | F1-score | 97.7 | 63.9 | 34.6 |
| MobileNetV2 | Accuracy | 99.1 | 74.6 | 24.7 |
| | Precision | 98.5 | 96.2 | 2.3 |
| | Recall | 99.6 | 47.6 | 52.2 |
| | F1-score | 99.0 | 63.7 | 35.7 |
| ResNet18 | Accuracy | **99.9** | 72.1 | 27.8 |
| | Precision | **99.9** | 97.9 | 2.0 |
| | Recall | **99.9** | 41.2 | 58.8 |
| | F1-score | **99.9** | 58.0 | 41.9 |
| Ours | Accuracy | 99.8 | **80.8** | 19.0 |
| | Precision | **99.9** | **98.8** | 1.1 |
| | Recall | 99.7 | **59.6** | **40.2** |
| | F1-score | 99.8 | **74.3** | 25.5 |

The results in Table II indicate that models trained on clean data can effectively distinguish between clean infant cries and noise on the clean testing dataset, achieving an accuracy score of 99.8% and an F1-score of 99.8%. However, when evaluated on the noisy testing dataset, performance metrics obtained by all methods deteriorate significantly. Especially, the recall scores drop the most significantly. This result shows that the models trained with only clean samples possess a low tolerance for noise and may not have adequately captured the unique patterns of infant cries. This result further implies the limited generalization capability of the models and suboptimal performance in practical applications when the models are trained with the clean samples only. In addition, the Accuracy and F1-score obtained by our method on the noisy testing data are 80.8% and 74.3%, respectively, which are higher than the counterparts of other methods. Therefore, its robustness to noise is superior to other methods.

Previous studies have shown that incorporating noisy data into the training process is essential for enhancing the generalization ability and robustness of a model [49]. In order to simulate the multi-source noise interference of home environments, we augment each training clip by introducing two to three noise segments and maintain single-channel. We dynamically vary SNRs from 0 to -20 dB for simulating scenarios in real-world environments where infant cries appear as both foreground and background sounds. We evaluate different methods on the testing sets with various SNRs (clean, 0, -10, and -20 dB). Experimental result are illustrated in Table III.

TABLE III EXPERIMENTAL RESULTS OBTAINED BY DIFFERENT METHODS WHEN THEIR MODELS ARE TRAINED USING NOISY DATASET AND TESTED ON NOISY DATASETS WITH DIFFERENT SNRs (IN %).

| Methods | Metric | Clean | 0 dB | -10 dB | -20 dB |
|---|---|---|---|---|---|
| BiLSTM | Accuracy | 89.5 | 85.3 | 81.8 | 72.3 |
| | Precision | 84.2 | 83.8 | 81.2 | 75.5 |
| | Recall | 95.6 | 85.0 | 79.7 | 60.3 |
| | F1-score | 89.5 | 84.4 | 80.4 | 67.0 |
| ECAPA-TDNN | Accuracy | 92.4 | 88.9 | 87.6 | 79.2 |
| | Precision | 90.4 | 88.1 | 88.0 | 85.0 |
| | Recall | 93.8 | 88.2 | 85.2 | 67.4 |
| | F1-score | 92.1 | 88.1 | 86.6 | 75.2 |
| MobileNetV2 | Accuracy | 96.6 | 94.0 | 87.1 | 79.2 |
| | Precision | 96.3 | 93.2 | 80.7 | 89.4 |
| | Recall | 96.5 | 94.1 | 95.3 | 63.1 |
| | F1-score | 96.4 | 93.6 | 87.4 | 74.0 |
| ResNet18 | Accuracy | 97.9 | 96.4 | 90.7 | 84.8 |
| | Precision | 97.7 | 96.8 | 91.7 | **95.7** |
| | Recall | 97.8 | 95.5 | 88.2 | 70.7 |
| | F1-score | 97.7 | 96.1 | 89.9 | 81.3 |
| Ours | Accuracy | **99.0** | **98.4** | **96.7** | **91.3** |
| | Precision | **98.4** | **97.9** | **97.1** | 90.4 |
| | Recall | **99.5** | **98.8** | **95.8** | **91.2** |
| | F1-score | **98.9** | **98.3** | **96.4** | **90.8** |

Under various SNRs, our method consistently achieves higher accuracies and F1-scores than the counterparts obtained by other methods. Due to the different feature patterns between the noisy training dataset and the clean training dataset, all methods achieve lower Accuracies and F1-scores (as shown in the "Clean" column of Table III) on the clean testing dataset compared to the counterparts in Table II (as shown in the "Clean" column of Table II). However, all methods achieve significantly higher Accuracies and F1-scores (as shown in the columns "0dB", "-10dB" and "-20dB" of Table III) compared to the counterparts in Table II (as shown in the column of "Noisy" of Table II) when all methods are evaluated on the noisy testing datasets. Furthermore, the

BiLSTM-based method performs the worst among all methods in terms of both Accuracy and F1-score, which is consistent with the result in [31]. This result highlights the challenges associated with modeling complex infant cries directly in the time domain, and the necessity of leveraging convolution operations for acquiring local information from the input Mel-spectrogram. In addition, when the SNRs decrease from 0 dB to -20 dB, the Accuracies and F1-scores of our method decrease by 7.1% (98.4% - 91.3%) and 7.5% (98.3% - 90.8%), respectively. However, the decreases of both Accuracy and F1-score of other methods are much greater than the counterparts obtained by our method. That is, compared to other methods, our method has higher robustness to noise.

An ablation experiment is done to assess the contribution of each main module of the proposed model. We use the same training configuration as done in the experiments of Table III and evaluate our method on the testing data with a fixed 0 dB SNR. As shown in Table IV, the method with all modules achieves the best overall performance. Notably, the ADM has the most significant impact on the performance of our method, underscoring the importance in enhancing robustness of our method to noise.

TABLE IV RESULTS OF ABLATION EXPERIMENT OBTAINED BY OUR METHOD WITH DIFFERENT MAIN MODULES (IN %).

| Methods | Accuracy | Precision | Recall | F1-score |
|---|---|---|---|---|
| *Without ADM* | 95.0 | 96.7 | 92.5 | 94.6 |
| *Without ESA* | 95.2 | 96.2 | 93.3 | 94.7 |
| *Without CCA* | 97.4 | 97.8 | 96.7 | 97.2 |
| *Without Multi-scale* | 96.0 | 95.6 | 95.8 | 95.7 |
| *With all modules* | **98.4** | **97.9** | **98.8** | **98.3** |

In addition to the comparison of Accuracy and F1-score, we compare our method with other methods in complexity. Specifically, Floating Point Operations (FLOPs) and Number of Parameters (NP) are used to measure the computational complexity and memory requirement of different methods, respectively. Table V presents the values of both NP and FLOPs of different methods. The values of NP and FLOPs of our method are 1.54 Million and 0.46 Giga, respectively. Among all methods, our method has the smallest NP value and the second smallest FLOPs value. That is, our method has advantage over other methods in memory requirement. It has lower computation complexity than other methods except the MobileNetV2-based method.

TABLE V COMPLEXITY COMPARISON OF DIFFERENT METHODS.

| Methods | NP (Million) | FLOPs (Giga) |
|---|---|---|
| *BiLSTM* | 6.61 | 1.31 |
| *ECAPA-TDNN* | 5.12 | 0.85 |
| *MobileNetV2* | 2.23 | **0.30** |
| *ResNet18* | 11.18 | 1.81 |
| *Ours* | **1.54** | 0.46 |

## V. CONCLUSIONS

In this paper, we proposed a method for infant cry detection in noisy environments. The model in the proposed method consists of blueprint separable convolutions with a time-frequency RNN-based adaptive denoising module and attention mechanisms tailored for both spatial and channel dimensions. By constructing a multi-source dataset and applying data augmentation, our method effectively captured the intricate acoustic patterns of infant cries and adaptively suppressed diverse noise interferences. Experimental results showed that our method exhibited superior robustness and accuracy across varying SNR conditions with lower complexity compared to the state-of-the-art methods. The above results not only validated the efficacy of our method but also highlighted its suitability for deployment of devices with limited computing resources. Future work will focus on deploying the proposed method to intelligent audio terminals and expanding the infant cry dataset to enhance the validity of the proposed method in even more challenging and diverse real-world environments.